\documentclass[aps,preprint,showpacs,nofootinbib]{revtex4}
\usepackage{amsfonts}
\usepackage{amsmath}
\usepackage{amssymb}
\usepackage{relsize}
\usepackage{graphicx}
\usepackage{pdfpages}
\usepackage{subfigure}
\usepackage{float}
\usepackage[caption = false]{subfig}
\usepackage{color}
\usepackage{multirow}
\begin{document}
%\addcolour{Mauve}{0.5}{0.0}{0.5}
%\includegraphics{sears1a.eps}

\title{Helium-like atoms. The Green's function approach to the Fock expansion calculations}

\author{Evgeny Z. Liverts}
\affiliation{Racah Institute of Physics, The Hebrew University, Jerusalem 91904,
Israel}

\author{Nir Barnea}
\affiliation{Racah Institute of Physics, The Hebrew University, Jerusalem 91904,
Israel}

%\lyxaddress{Racah Institute of Physics, The Hebrew University, Jerusalem 91904,Israel}
\begin{abstract}

The renewed Green's function approach to calculating the angular Fock coefficients, $\psi_{k,p}(\alpha,\theta)$  is presented.
The final formulas are simplified and specified to be applicable for analytical as well as numerical calculations.
%The hyperspherical angles $\theta=0,\pi$ (arbitrary $\alpha$) or $\alpha=0,\pi$ (arbitrary $\theta$) are indicated as corresponding to the Green's function formulas possessing the physical meaning.
The Green's function formulas with the hyperspherical angles $\theta=0,\pi$ (arbitrary $\alpha$) or $\alpha=0,\pi$ (arbitrary $\theta$) are indicated as corresponding to the angular Fock coefficients possessing physical meaning.
The most interesting case of $\theta=0$ corresponding to a collinear arrangement of the particles is studied in detail.
It is emphasized that this case represents the generalization of the specific cases of the electron-nucleus ($\alpha=0$) and electron-electron ($\alpha=\pi/2$) coalescences.
It is shown that the Green's function method for $\theta=0$ enables us to calculate
any component/subcomponent of the angular Fock coefficient
in the form of a single series representation with arbitrary angle $\theta$.
Those cases, where the Green's function approach can not be applied, are thoroughly studied, and the corresponding solutions are found.

\end{abstract}


\maketitle

\section{Introduction}\label{S0}

The helium-like atoms represent the two-electron atomic systems which can serve as an excellent model for testing the new quantum theories.
It is therefore not surprising that a large number of scientific studies were devoted to solving the problem of theoretical calculations of the electronic structure of these quantum systems.
As far back as the 1935s Bartlett et al. \cite{B35} showed that no ascending power series in the interparticle coordinates $r_1,~r_2$ and $r_{12}$ can be a formal solution of the Schr\"{o}dinger equation
for the $~^1S$-state of helium.
A rigorous investigation of the helium atom has been made by Fock \cite{FOCK} who proposed the following expansion of the corresponding wave function in the vicinity of the triple coalescence (collision) point (when $r_1=r_2=0$)
\begin{equation}\label{I1}
\Psi(r,\alpha,\theta)=\sum_{k=0}^{\infty}r^k\sum_{p=0}^{[k/2]}\psi_{k,p}(\alpha,\theta)(\ln r)^p .
\end{equation}
The hyperspherical coordinates $r=\sqrt{r_1^2+r_2^2}$, and
\begin{equation}\label{I2}
\alpha=2\arctan \left(r_2/r_1\right),~~~\theta=\arccos\left[(r_1^2+r_2^2-r_{12}^2)/(2r_1r_2)\right]
\end{equation}
were used. The important feature of the Fock expansion (\ref{I1}) is including the logarithmic function of the hyperspherical radius $r$.
Leray \cite{LER} and Morgan \cite{MORG} have rigorously justified the necessity of this expansion.
Ermolaev and Demkov \cite{DEM,ES1} studied the extension of the Fock expansion to an arbitrary system of charged particles and to state of any symmetry.
Ermolaev \cite{ES2} and David \cite{DAV} extended the work of Fock obtaining $\psi_{2,0}$ by expansion into the hyperspherical harmonics.
Feagin et al. \cite{FEM} presented a procedure for matching the series representation (\ref{I1}) onto the hyperspherical adiabatic channel representation.
Pluvinage \cite{PLUV} somewhat generalized the Fock expansion to be applicable to any $S$ state, and its first two terms were determined.
The angular Fock coefficients, $\psi_{k,p}(\alpha,\theta)$ have been calculated up to $k=2$ by Forrey \cite{FOR} using the hyperspherical angles that differ from the angles defined by Eq.(\ref{I2}).
The most comprehensive investigation of the methods of calculation of the angular Fock coefficients (AFC)
was presented in the works of Abbott, Gottschalk and Maslen \cite{AM1,GAM2,GM3}.

Using the original hyperspherical angles $\alpha$ and $\theta$, methods of calculation of the AFCs have been supplemented and extended in the works \cite{LEZ1,LEZ2,LEZ3}.
It was proposed \cite{LEZ1} to separate the AFCs into the components in the following manner
\begin{equation}\label{I3}
\psi_{k,p}(\alpha,\theta)=\sum_{j=p}^{k-p} \psi_{k,p}^{(j)}(\alpha,\theta) Z^j.
\end{equation}
 The component $\psi_{k,p}^{(j)}(\alpha,\theta)$ being associated with a definite power of the nucleus charge $Z$ of the two-electron atomic system with energy $E$ and infinite nuclear mass, satisfies the individual Fock recurrence relations (IFRR)
\begin{subequations}\label{I4}
\begin{align}
\left[ \Lambda^2-k(k+4)\right]\psi_{k,p}^{(j)}\left(\alpha,\theta\right)=h_{k,p}^{(j)}\left(\alpha,\theta\right),
~~~~~~~~~~~~~~~~~~~~~~~~~~~~~~~~~~~~~~~~~~~~~~~~~~~\label{I4a}\\
h_{k,p}^{(j)}=2(k+2)(p+1)\psi_{k,p+1}^{(j)}+(p+1)(p+2)\psi_{k,p+2}^{(j)}-2 V_0 \psi_{k-1,p}^{(j)}-2 V_1 \psi_{k-1,p}^{(j-1)}+2 E \psi_{k-2,p}^{(j)},\label{I4b}~~~
\end{align}
\end{subequations}
where the hyperspherical angular momentum operator, projected on $S$ states, is defined as
\begin{equation}\label{I5}
\Lambda^2=-4\left[\frac{\partial^2}{\partial\alpha^2}+2\cot\alpha\frac{\partial}{\partial\alpha}+ \frac{1}{\sin^2\alpha}\left( \frac{\partial^2}{\partial \theta^2}+\cot \theta \frac{\partial}{\partial\theta}\right)\right].
\end{equation}
The dimensionless Coulomb potentials representing the electron-electron and electron-nucleus interactions, respectively, are
\begin{equation}\label{I6}
V_0\equiv \frac{r}{r_{12}}=\frac{1}{\sqrt{1-\sin \alpha \cos \theta}},~~~~~~~~~~
V_1\equiv -\frac{r}{r_1}-\frac{r}{r_2}=-\frac{2\sqrt{1+\sin \alpha}}{\sin \alpha}.
\end{equation}
It was shown in the works \cite{LEZ1,LEZ2,LEZ3} that any component of the AFC can be represented by a single series of the form
\begin{equation}\label{I7}
\psi_{k,p}^{(j)}(\alpha,\theta)=\sum_{l=0}^\infty \omega_l^{(k,p;j)}(\alpha)(\sin \alpha)^lP_l(\cos\theta),
\end{equation}
where some of the functions $\omega_l^{(k,p;j)}$ can be constant including zero. $P_l(z)$ are the Legendre polynomials.
%--------------------------------------------------------------------------------------------------------------

Remind that in his original work \cite{FOCK} Fock used the Green's function approach to solve the inhomogeneous differential equations similar to Eq.(\ref{I4}).
Subsequent authors, as a rule, employed  another methods.
The powerful method using the Green's functions was undeservedly forgotten, probably because of some complexity and ambiguity of its application.

In this article we propose to return to the Green's function method for solving the IFRR (\ref{I4}).
We simplify and concretize the final formulas what enables us to calculate the AFC components/subcomponents both analytically and numerically.
This approach can be particularly successful in calculating the collinear helium-like atomic systems (when the hyperspherical angle $\theta=0$), and hence, in calculating the specific cases of the electron-electron ($\alpha=\pi/2$) and electron-nucleus ($\alpha=0$) coalescences.

\section{Green's function solutions}\label{S1}

Using the generalized Green's function approach it was shown \cite{ABT} that any AFC can be calculated by the following relationships:
\begin{equation}\label{K1}
\psi_{k,p}(\alpha',\theta')=\frac{1}{8\pi^2}\int h_{k,p}(\alpha,\theta)\frac{\cos\left[\left(\frac{k}{2}+1\right)\gamma\right]}
{\sin \gamma} d \Omega,~~~~~~~~~~~~~~~~~~~~~~~~~~~~k~~~odd
\end{equation}
\begin{equation}\label{K2}
\psi_{k,p}(\alpha',\theta')=\sum_{l=0}^{k/2}A_{kl}Y_{kl}(\alpha',\theta')+
\frac{1}{8\pi^3}\int h_{k,p}(\alpha,\theta)(\pi-\gamma)\frac{\cos\left[\left(\frac{k}{2}+1\right)\gamma\right]}
{\sin \gamma} d \Omega,~~~~~~k~~~even
\end{equation}
where the volume element is
\begin{equation}\label{K3}
d\Omega=\pi^2 \sin^2\alpha \sin\theta d\alpha d\theta,~~~~~~~~~~~~~~~~~\alpha\in [0,\pi],~\theta\in [0,\pi]
\end{equation}
and the unnormalized hyperspherical harmonics (HHs) are
\begin{equation}\label{K4}
Y_{kl}(\alpha,\theta)=(\sin \alpha)^l~C_{k/2-l}^{(l+1)}(\cos\alpha)P_l(\cos\theta),~~~~~~~~~~k=0,2,4,...; l=0,1,2,...,k/2 .
\end{equation}
Notation $C_n^{(m)}(x)$ corresponds to the Gegenbauer polynomials.

The angle $\gamma$ was defined as \cite{ABT}
\begin{equation}\label{K5}
\cos \gamma =\cos \alpha \cos \alpha'+\sin \alpha \sin \alpha'\cos \theta \cos \theta'.
\end{equation}

Remind that in the original paper of Fock \cite{FOCK} the formulas that are similar to (\ref{K1}) and (\ref{K2}) were presented as well.
The angle corresponding to $\gamma$ was there denoted as $\omega$ and was defined \cite{FOCK} as
\begin{equation}\label{K6}
\cos \omega =\cos \alpha \cos \alpha'+\sin \alpha \sin \alpha'
\left[\cos \theta \cos \theta'+\sin \theta \sin \theta' \cos(\varphi-\varphi')\right].
\end{equation}
The very important remark of Fock \cite{FOCK} was the following : "Only the solutions, which are not dependent on the angle $\varphi$, are of the physical meaning".
It is seen from definition (\ref{K6}) that the angle $\omega$ becomes independent on the angle $\varphi$ just for $\theta'=0,\pi$ (arbitrary $\alpha'$) or for $\alpha'=0,\pi$ (arbitrary $\theta'$) .
Note that for these values of the angles $\alpha'$ and $\theta'$, the angle $\omega$ defined by Eq.(\ref{K6}) and the angle $\gamma$ defined by Eq.(\ref{K5}) are coincident.

On examples of all the AFCs (or their components/subcomponents) presented in Refs.\cite{LEZ1,LEZ2,LEZ3}, we have verified \emph{numerically} (by Wolfram \emph{Mathematica}) that formulas (\ref{K1}) and (\ref{K2}) are correct namely for the angles $\alpha'$ and $\theta'$ mentioned above.
There are exceptions that will be discussed in what follows.
We are interested, first of all, in the angle $\theta'=0$. The reasons will become clear from what follows.
It would be useful to note that the angle  $\theta'=0$ corresponding to a collinear arrangement of the particles enables us to calculate the AFCs at the two-particle atomic coalescences corresponding to $\alpha'=0$ and $\alpha'=\pi/2$.
Thus, formulas (\ref{K1}) and (\ref{K2}) being suitable for the components/subcomponents of the AFCs considered for the angles mentioned above, become
\begin{equation}\label{K7}
\psi_{k,p}^{(j)}(\alpha',0)=\frac{1}{8}\int_0^{\pi}d\alpha \sin^2\alpha \int_0^{\pi}d\theta \sin \theta~ h_{k,p}^{(j)}(\alpha,\theta)\frac{\zeta(\gamma)\cos\left[\left(\frac{k}{2}+1\right)\gamma\right]}
{\sin \gamma},
\end{equation}
where
\begin{equation}\label{K8}
\zeta(\gamma)=
\left\{ \begin{array}{c}
\mathlarger{1}~~~~~~~~~~~~~~~~~k~~~odd\\
\mathlarger{1-\gamma/\pi}~~~~~~~~~k~~~even\\
\end{array}\right.
\end{equation}
The redefined angle $\gamma$ now reads
\begin{equation}\label{K9}
\cos \gamma =\cos \alpha \cos \alpha'+\sin \alpha \sin \alpha'\cos \theta.
\end{equation}
It should be noticed that the Green's function formulas for the AFCs
%(\ref{K1}) and (\ref{K2})
follow from the addition theorem
\begin{equation}\label{K9a}
\sum_{l=0}^{k/2}Y_{kl}(\alpha,\theta)Y_{kl}(\alpha',0)=\frac{k+2}{2\pi^3}~C_{k/2}^{(1)}(\cos \gamma)
\end{equation}
for HHs, where the angle $\gamma$ is defined by Eq.(\ref{K9})  (see Ref.\cite{KLAR}).
The relationship (\ref{K9a}) confirms the correctness of the result (\ref{K7})-(\ref{K9}).

One should emphasize that according to the basic definition (\ref{K2}), Eq.(\ref{K7}) with even $k$ enables us to obtain the "pure" component/subcomponent (see Ref.\cite{LEZ1}) directly, that is without any more operations discussed in Ref.\cite{LEZ1,LEZ2,LEZ3}.

Moreover, it is very important that Eq.(\ref{K7}) enables us to obtain the $\alpha$-dependent functions of the general expansion (\ref{I7}) in the form
\begin{equation}\label{K10}
\omega_l^{(k,p;j)}(\alpha')=\frac{1}{8(\sin \alpha')^l}\int_0^{\pi}d\alpha (\sin\alpha)^{l+2} h_l(\alpha)
\int_0^{\pi} \frac{\zeta(\gamma)\cos\left[\left(\frac{k}{2}+1\right)\gamma\right]}{\sin \gamma}
P_l(\cos \theta)\sin \theta~d\theta,
\end{equation}
where $h_l(\alpha)\equiv h_l^{(k,p;j)}(\alpha)$ is defined by the following expansion of the rhs (\ref{I4b}) of the IFRR (\ref{I4a})
\begin{equation}\label{K11}
h_{k,p}^{(j)}(\alpha,\theta)=\sum_{l=0}^\infty h_l(\alpha)(\sin \alpha)^lP_l(\cos\theta),
\end{equation}
which are similar to expansion (\ref{I7}).
To derive representation (\ref{K10}) we used $P_l(1)=1$.

\section{Solutions for exceptions}\label{S2}

It was mentioned in the previous section that there are components/subcomponents which cannot be found by the formulas (\ref{K7})-(\ref{K9}) or (\ref{K10}). It is clear that, first of all, this concerns the IFRRs (\ref{I4}) with zero rhs $h_{k,p}^{(j)}$ corresponding to $p=k/2$ for even $k$. However, a simple solution to this problem was presented in Ref.\cite{AM1}.
Then, it should be noted that the numerical calculations of the subcomponents $\psi_{4,1}^{(2b)}$ and $\psi_{4,1}^{(2d)}$ \cite{LEZ1} show that the results of the Green's function approach and the analytic results presented in Ref.\cite{LEZ1,LEZ2} are not coincident.
In particular, for subcomponent $\psi_{4,1}^{(2d)}$ there are discrepancies in the expansion terms corresponding to $l=0$ and $l=2$ (see Appendix C and Eq.(93) in Ref.\cite{LEZ1}).
We shall also obtain a similar discrepancy, for example, in case of solving the IFFR
\begin{equation}\label{P1}
\left( \Lambda^2-12\right)\psi_{2,0}^{(1x)}=h_{2,0}^{(1x)},
\end{equation}
with the rhs (see Eq.(50) of Ref.\cite{LEZ1})
\begin{equation}\label{P2}
h_{2,0}^{(1x)}\equiv 8\psi_{2,1}=-\frac{8(\pi-2)}{3\pi}\sin \alpha \cos \theta.
\end{equation}
The reason for such problems is the following.
Even in the Fock's paper \cite{FOCK} it was underlined that for even $k$ (corresponding to integer $n$ in terms of the original paper) the solution of the IFRR (\ref{I4}) can be found in the Green's function form only if the rhs of this equation obey the $k/2 +1$ orthogonality conditions
\begin{equation}\label{P3}
\int_0^\pi d\alpha \sin^2\alpha \int_0^\pi d\theta \sin \theta~h_{k,p}^{(j)}(\alpha,\theta)Y_{kl}(\alpha,\theta)=0,
\end{equation}
that is when the rhs of the IFRR (\ref{I4}) is orthogonal to solutions of the homogeneous equation associated with it.

It follows from condition (\ref{P3}) that the Green's function approach cannot be used to calculate the solution of the IFRR (\ref{I4}) with the rhs $h_{k,l}^{(j)}$ which is proportional to one of the HHs, $Y_{kl}(\alpha,\theta)~~~(l=0,1,...k/2)$.
We come across exactly such cases when we are going to calculate $\psi_{2,0}^{(1x)}$ with $h_{2,0}^{(1x)}\varpropto Y_{21}(\alpha,\theta)$ or  $\psi_{4,1}^{(2b)}$ with $h_{4,1}^{(2b)}\varpropto Y_{40}(\alpha,\theta)$.

In case of the rhs $h_{k,p}^{(J)}$ is not proportional to $Y_{kl}$, but doesn't satisfy the orthogonality conditions (\ref{P3}), the Green's function formula (\ref{K7}) still can be applied
to find the solution $\psi_{k,p}^{(J)}(\alpha,\theta)$ to the corresponding IFRR.
The superscript $J$ of the rhs denotes any subcomponent or any term of expansion (\ref{K11}).
The aforementioned terms of expansion (\ref{K11}) for subcomponent $\psi_{4,1}^{((2d)}$ are the examples.
So, to solve the problem we propose to apply the violent orthogonalization.
This means that one should replace the considered rhs $h_{k,p}^{(J)}$ by
\begin{equation}\label{P4}
\widetilde{h}_{k,p}^{(J)}(\alpha,\theta)=h_{k,p}^{(J)}(\alpha,\theta)+\sum_{l=0}^{k/2}C_l Y_{kl}(\alpha,\theta),
\end{equation}
where (denoting $C_l\equiv C_l^{(k,p;J)}$ for simplicity)
\begin{equation}\label{P5}
C_l=-\int_0^\pi d\alpha \sin^2\alpha \int_0^\pi d\theta \sin \theta~h_{k,p}^{(J)}(\alpha,\theta)Y_{kl}(\alpha,\theta)
\left[\int_0^\pi d\alpha \sin^2\alpha \int_0^\pi d\theta \sin \theta~Y_{kl}^2(\alpha,\theta)\right]^{-1}.
\end{equation}
Note that, as a rule, only one coefficient $C_l$ should be calculated.
The resultant IFRR
\begin{equation}\label{P6}
\left[ \Lambda^2-k(k+4)\right]\widetilde{\psi}_{k,p}^{(J)}\left(\alpha,\theta\right)=\widetilde{h}_{k,p}^{(J)}\left(\alpha,\theta\right)
\end{equation}
can be solved by the Green's function methods described above.
It is clear that the required solution is of the form
\begin{equation}\label{P7}
\psi_{k,p}^{(J)}\left(\alpha,\theta\right)=\widetilde{\psi}_{k,p}^{(J)}\left(\alpha,\theta\right)-\sum_{l=0}^{k/2}C_l F_{k,l}\left(\alpha,\theta\right),
\end{equation}
where the function $F_{k,l}(\alpha,\theta)$ represents the "pure" physical solution of the equation
\begin{equation}\label{P8}
\left[ \Lambda^2-k(k+4)\right]F_{k,l}\left(\alpha,\theta\right)=Y_{k,l}\left(\alpha,\theta\right).
\end{equation}
Fortunately, such a solution can be obtained in general analytic form (see also Ref.\cite{GM3}).
To solve Eq.(\ref{P8}), first of all, it is convenient to present the solution in the form
\begin{equation}\label{P9}
F_{k,l}\left(\alpha,\theta\right)=f_{k,l}(\alpha)(\sin \alpha)^{-1}P_l(\cos \theta),
\end{equation}
where definition (\ref{K4}) of HH, and the relation (39) from Ref.\cite{LEZ1} were used.
This substitution enables one to reduce Eq.(\ref{P8}) to the following equation of one variable
\begin{equation}\label{P10}
\frac{d^2}{d\alpha^2}f_{k,l}(\alpha)+\left[\left(\frac{k}{2}+1\right)^2-\frac{l(l+1)}{\sin^2\alpha}\right]f_{k,l}(\alpha)=-\frac{1}{4}h_{k,l}(\alpha),
\end{equation}
where
\begin{equation}\label{P11}
h_{k,l}(\alpha)=(\sin \alpha)^{l+1}C_{k/2-l}^{(l+1)}(\cos \alpha)=-\frac{2^{-l+\frac{1}{2}}}{\sqrt{\pi}~l!}(-1)^l\sqrt{\sin \alpha}~Q_{\frac{k+1}{2}}^{l+\frac{1}{2}}(\cos \alpha).
\end{equation}
The representations (\ref{P11}) follow from definition (\ref{K4}) of the HH, and from the relationships between the Gegenbauer polynomials and the Legendre functions (see, e.g., \cite{BAT}).
The fundamental solutions of the homogeneous equation associated with Eq.(\ref{P10}) are
\begin{equation}\label{P12}
u_l(\alpha)\equiv u_{k,l}(\alpha)=\sqrt{\sin \alpha}~P_{\frac{k+1}{2}}^{l+\frac{1}{2}}(\cos \alpha),
\end{equation}
\begin{equation}\label{P13}
v_l(\alpha)\equiv v_{k,l}(\alpha)=\sqrt{\sin \alpha}~Q_{\frac{k+1}{2}}^{l+\frac{1}{2}}(\cos \alpha),
\end{equation}
where $P_\nu^\mu(z)$ and $Q_\nu^\mu(z)$ are the associated Legendre functions of the first and second kinds, respectively.
The subscript $k$ of the homogeneous solutions is omitted for the sake of simplicity.
It can be shown that the Wronskian for the homogeneous functions (\ref{P12}) and (\ref{P13}) is a rational number:
\begin{equation}\label{P14}
W_{k,l}\equiv v_l(\alpha)\frac{d}{d\alpha}u_l(\alpha)-u_l(\alpha)\frac{d}{d\alpha}v_l(\alpha)=\Gamma\left(\frac{k}{2}+l+2\right)\left/\Gamma\left(\frac{k}{2}-l+1\right)\right. ,
\end{equation}
where $\Gamma(z)$ is the Euler gamma function.

It is well known that the general solution of the inhomogeneous equation (\ref{P10}) can be written as follows
\begin{equation}\label{P15}
f_{k,l}(\alpha)=f_{k,l}^{(p)}(\alpha)+A_l^{(u)}u_l(\alpha)+A_l^{(v)}v_l(\alpha).
\end{equation}
The \textbf{particular} solution of Eq.(\ref{P10}) can be found by the method of variation of parameters in the form
\begin{equation}\label{P16}
f_{k,l}^{(p)}(\alpha)=\frac{\Gamma\left(\frac{k}{2}-l+1 \right)2^{-l-\frac{3}{2}}(-1)^l}{\Gamma\left(\frac{k}{2}+l+2\right)\sqrt{\pi}~l!}
\sqrt{\sin \alpha}\left[P_{\frac{k+1}{2}}^{l+\frac{1}{2}}(\cos \alpha)U_{k,l}(\alpha)-Q_{\frac{k+1}{2}}^{l+\frac{1}{2}}(\cos \alpha)V_{k,l}(\alpha)
\right],
\end{equation}
where
\begin{equation}\label{P17}
U_{k,l}(\alpha)=\int_0^\alpha Q_{\frac{k+1}{2}}^{l+\frac{1}{2}}(\cos \beta)Q_{\frac{k+1}{2}}^{l+\frac{1}{2}}(\cos \beta)\sin \beta~d\beta,
\end{equation}
\begin{equation}\label{P18}
V_{k,l}(\alpha)=\int_0^\alpha P_{\frac{k+1}{2}}^{l+\frac{1}{2}}(\cos \beta)Q_{\frac{k+1}{2}}^{l+\frac{1}{2}}(\cos \beta)\sin \beta~d\beta.
\end{equation}
The Wronskian (\ref{P14}) was certainly used to obtain the representation (\ref{P16})-(\ref{P18}),
which is valid not only for even $k$ (see definition (\ref{K4}) of HHs) but for any integer $k\geq0$.

It can be shown that the leading terms of series expansions of the homogeneous solutions (\ref{P12}), (\ref{P13}) and the particular solution (\ref{P16})-(\ref{P18}) are proportional to:
\begin{equation}\label{P19}
u_l(\alpha)\underset{\alpha\rightarrow 0}{\propto} \alpha^{-l},~~~~~~~~~~~~~~~~u_l(\alpha)\underset{\alpha\rightarrow \pi}{\propto} (\alpha-\pi)^{-l},
\end{equation}
\begin{equation}\label{P20}
v_l(\alpha)\underset{\alpha\rightarrow 0}{\propto} \alpha^{l+1},~~~~~~~~~~~~~~~~v_l(\alpha)\underset{\alpha\rightarrow \pi}{\propto} (\alpha-\pi)^{l+1},
\end{equation}
\begin{equation}\label{P21}
f_{k,l}^{(p)}(\alpha)\underset{\alpha\rightarrow 0}{\propto} \alpha^{l+3},~~~~~~~~~~~~~~~~f_{k,l}^{(p)}(\alpha)\underset{\alpha\rightarrow \pi}{\propto} (\alpha-\pi)^{-l}.
\end{equation}
It is seen that for any $l\geq0$ the particular solution
\begin{equation}\label{P22}
F_{k,l}^{(p)}(\alpha,\theta)=f_{k,l}^{(p)}(\alpha)(\sin \alpha)^{-1} P_l(\cos \theta)
\end{equation}
of Eq.(\ref{P8}) is finite at the point $\alpha=0$ (arbitrary $\theta$), and it is singular at the point $\alpha=\pi $ (arbitrary $\theta$),
whereas the homogeneous solution $u_l(\alpha)$ is singular at the both $\alpha$-points mentioned above.
This means that the physical solution of Eq.(\ref{P8}) can be obtained in the form
\begin{equation}\label{P23}
F_{k,l}(\alpha,\theta)=\frac{P_l(\cos \theta)}{\sin \alpha}
\left\{ \begin{array}{c}
\mathlarger{g_{k,l}(\alpha)}~~~~~~~~~~~~~~~~~0\leq\alpha\leq\pi/2~\\
\mathlarger{\pm g_{k,l}(\pi-\alpha)}~~~~~~~~~\pi/2\leq\alpha\leq\pi~\\
\end{array}\right.,
\end{equation}
where one should set $A_l^{(u)}=0$ to obtain
\begin{equation}\label{P24}
g_{k,l}(\alpha)=f_{k,l}^{(p)}(\alpha)+A_l^{(v)}v_l(\alpha).
\end{equation}
Note that the plus and minus signs in Eq.(\ref{P23}) correspond to the symmetric and antisymmetric physical solutions, respectively.
Moreover, the physical solution must preserve (as a rule) the continuity property over the whole range under consideration. This means that at the symmetric (antisymmetric) point $\alpha=\pi/2$ one should set
\begin{equation}\label{P25}
\left.\frac{d}{d \alpha}g_{k,l}(\alpha)\right|_{\alpha=\pi/2}=0 .
\end{equation}
In fact, condition (\ref{P25}) corresponds to the standard matching of the logarithmic derivatives at $\alpha=\pi/2$.
Note that one should certainly require the continuity of the function $g_{k,l}(\alpha)/\sin \alpha$, however, for the point $\alpha=\pi/2$ the result will be the same as under condition (\ref{P25}).
Application of this condition enables us to calculate the coefficient $A_l^{(v)}$ by the formula
\begin{equation}\label{P26}
\left.\left.A_l\equiv A_l^{(v)}=-\left(\frac{d f_{k,l}^{(p)}(\alpha)}{d\alpha}\right/\frac{d v_{k,l}(\alpha)}{d\alpha}\right)\right|_{\alpha=\pi/2} .
\end{equation}
It can be easily shown that
\begin{equation}\label{P27}
\left.\frac{d v_{k,l}(\alpha)}{d\alpha}\right|_{\alpha=\pi/2}=
\frac{\sqrt{\pi}~2^{l+\frac{1}{2}}\Gamma\left(\frac{k+2l+6}{4}\right)}{\Gamma\left(\frac{k-2l+2}{4}\right)}
\sin\left(\frac{\pi(k+2l)}{4}\right).
\end{equation}
Whence, the derivative  $d v_{k,l}(\alpha)/d\alpha$ at $\alpha=\pi/2$ equals zero for \textbf{even} values of $(k+2l)/2$.
And hence, we can find the solution (\ref{P23}) possessing the continues logarithmic derivative at the point $\alpha=\pi/2$ only for \textbf{odd} values of $(k+2l)/2$ .
It can be shown that the property of "purity" of the physical solution of Eq.(\ref{P8}) follows from the property of continuity of the logarithmic derivative of $g_{k,l}(\alpha)$ at the point $\alpha=\pi/2$ .
Thus, using Eqs.(\ref{P24})-(\ref{P26}) for \textbf{odd} values of $(k+2l)/2$  we can derive the physical solution $F_{k,l}(\alpha,\theta)$ which possesses both the continuity and "purity" properties, simultaneously.

Remind  that the "pure" physical solution (see Ref.\cite{LEZ1}) of the form
\begin{equation}\label{P28}
F_{k,l}(\alpha,\theta)=F_{k,l}^{(p)}(\alpha,\theta)+B_l Y_{kl}(\alpha,\theta)
\end{equation}
must satisfy the equation
\begin{equation}\label{P29}
\int_0^\pi d \alpha \sin^2\alpha \int_0^\pi
F_{k,l}(\alpha,\theta)Y_{kl}(\alpha,\theta)\sin \theta~d\theta=0,
\end{equation}
which enables one to calculate the coefficient $B_l$.

As it was mentioned above, for even values of $(k+2l)/2$  we cannot find the physical solution of Eq.(\ref{P8}) possessing the continues logarithmic derivative at the point $\alpha=\pi/2$.
However, using Eqs.(\ref{P28})-(\ref{P29}) for this case, we can obtain the "pure" physical solution of Eq.(\ref{P8}).

According to representation (\ref{P23}), any physical solution $F_{k,l}(\alpha,\theta)$ to Eq.(\ref{P8}) is defined by the physical solution $g_{k,l}(\alpha)$ to Eq.(\ref{P10}).
The particular solution of Eq.(\ref{P10}) represented by Eq.(\ref{P16})-(\ref{P18}) can be derived in explicit form at least for given value of $l$ (see the Appendix). Using the technique described in this Section, it is possible to obtain the physical
% continues (and hence, ``pure")
solution of Eq.(\ref{P10}) in the form
%Such a solution for $l=0$ was derived in the most simple form
\begin{equation}\label{P30}
g_{k,l}(\alpha)=G_l\left(\frac{k}{2}+1,\alpha\right),
\end{equation}
where
\begin{equation}\label{P31}
G_l(n,\alpha)=\frac{1}{2^{l+4}~l! \sin^l \alpha}\left[\sin(n \alpha)S_{l,n}(\alpha)  +\cos(n \alpha)T_{l,n}(\alpha)  \right],
\end{equation}
which is valid only for even values of $k$ (unlike the particular solution) corresponding to the principal definition (\ref{K4}) of the HHs.
It is worth noting that the Green's function solution (\ref{K7}) contains the same angular factor $n=k/2+1$.

For $l=0$ one obtains the most simple solution of the form
\begin{equation}\label{P32}
G_0(n,\alpha)=\frac{1}{16n^2}
\left\{ \begin{array}{c}
\mathlarger{2 n \alpha \cos(n \alpha)-2\sin(n \alpha)}~~~~~~~~~~~~~~~~k=2,6,10,...~~~~~\\
\mathlarger{2 n \alpha \cos(n \alpha)-\sin(n \alpha)}~~~~~~~~~~~~~~~~~~k=0,4,8,12~,...~\\
\end{array}\right..
\end{equation}
In this case the particular solution calculated by Eqs.(\ref{P16})-(\ref{P18}) coincides with the continues physical solution corresponding to the odd values of $k/2$.
For even values of $k/2$, one obtains the "pure" physical solution (\ref{P32}) calculated by Eqs.(\ref{P28})-(\ref{P29}).
Functions $S_{l,n}(\alpha)$ and $T_{l,n}(\alpha)$ contained in representation (\ref{P31}) are presented in Tables \ref{T1} and \ref{T2} for $l=1,2,3$. The details of the corresponding derivations can be found in the Appendix.

\section{Conclusions} \label{S3}

%The important subject of the electronic structure of the helium-like atomic systems considered in Ref.\cite{LEZ1,LEZ2,LEZ3}
The development of an important topic of the electronic structure of the helium-like atomic systems considered in Refs.\cite{LEZ1,LEZ2,LEZ3} was continued.
The Green's function approach to calculating the angular Fock coefficients, $\psi_{k,p}(\alpha,\theta)$  has been revived.
The final formulas has been simplified and specified to be applicable for analytical as well as numerical calculations.
%It was established that only the hyperspherical angles $\theta=0,\pi$ (arbitrary $\alpha$) or $\alpha=0,\pi$ (arbitrary $\theta$) correspond to the Green's function formulas possessing the physical meaning.
It was established that only the Green's function formulas (\ref{K7})-(\ref{K9}) with the hyperspherical angles $\theta=0,\pi$ (arbitrary $\alpha$) or $\alpha=0,\pi$ (arbitrary $\theta$) describe the angular Fock coefficients possessing the physical meaning.
The most interesting case of $\theta=0$ corresponding to a collinear arrangement of the particles has been studied in detail.
It should be emphasized that this case represents the generalization of the cases of the electron-nucleus ($\alpha=0$) and electron-electron ($\alpha=\pi/2$) coalescences.
It is very important that the Green's function approach for $\theta=0$ enables us to calculate a single series representation (\ref{I7}) for any component/subcomponent of the angular Fock coefficient with arbitrary angle $\theta$.
Those cases, where the Green's function formulas can not be applied, were thoroughly studied, and the corresponding solutions were found.
The details and examples can be found in the Appendix.
The Wolfram Mathematica was used intensively.

\section{Acknowledgment}

%The author acknowledges Prof. Nir Barnea for useful discussions.
The author is grateful to Prof. Paul Abbott for providing his thesis.
%The author thanks to the anonymous referee for improving the presentation of his results in this article.
This work was supported by the PAZY Foundation.

\appendix

\section{}\label{SA}

To derive the solution of Eq.(\ref{P10}) it is convenient to use the following representations
\begin{equation}\label{A1}
P_\frac{k+1}{2}^\frac{1}{2}(\cos \alpha)=\sqrt{\frac{2}{\pi \sin \alpha}}\cos\left[\left(\frac{k}{2}+1\right)\alpha\right],~~
Q_\frac{k+1}{2}^\frac{1}{2}(\cos \alpha)=-\sqrt{\frac{\pi}{2 \sin \alpha}}\sin\left[\left(\frac{k}{2}+1\right)\alpha\right].
\end{equation}
which can obtained using the relations $(3.6.1.12)$\cite{BAT} and applying De Moivre's formula.
The associated Legendre functions (\ref{A1}) correspond to the homogeneous solutions (\ref{P12}) and (\ref{P13}) with $l=0$.
To obtain the Legendre functions with higher superscript the recurrence identity
\begin{equation}\label{A2}
R_\frac{k+1}{2}^{l+\frac{1}{2}}(\cos \alpha)=(1-2l)\cot \alpha R_\frac{k+1}{2}^{l-\frac{1}{2}}(\cos \alpha)-
\frac{1}{4}\left[k^2+4k-4l(l-2) \right] R_\frac{k+1}{2}^{l-\frac{3}{2}}(\cos \alpha),
\end{equation}
 together with the relations
\begin{equation}\label{A3}
P_\frac{k+1}{2}^{-\frac{1}{2}}(\cos \alpha)=-\frac{4}{\pi(k+2)}Q_\frac{k+1}{2}^\frac{1}{2}(\cos \alpha),~~~~
Q_\frac{k+1}{2}^{-\frac{1}{2}}(\cos \alpha)=\frac{\pi}{k+2}P_\frac{k+1}{2}^\frac{1}{2}(\cos \alpha)
\end{equation}
can be applied. $R_\nu^\mu(z)$ denotes the associated Legendre functions of the first or second kinds.
Inserting representations (\ref{A1}) into the general formulas (\ref{P16})-(\ref{P18}) with $l=0$, and performing integration, one obtains the particular solution of Eq.(\ref{P10}) in the form
\begin{equation}\label{A4}
f_{k,0}^{(p)}(\alpha)=\frac{1}{4(k+2)^2}\left\{(k+2)\alpha \cos\left[\left(\frac{k}{2}+1\right)\alpha\right]-
2 \sin\left[\left(\frac{k}{2}+1\right)\alpha\right] \right\}.
\end{equation}
Taking into account that the particular solution (\ref{A4}) just satisfies the continuity property (\ref{P25}), and hence the "purity" property, one obtains the physical solution (\ref{P32}) for $k=2,6,10,...$
As it was discussed in Sec.\ref{S2}, we cannot use Eqs.(\ref{P24})-(\ref{P26}) to derive the expression for the continues (at $\alpha=\pi/2$) physical solution for $k=0,4,8,..$, because of the factor $\sin[\pi(k+2l)/4]$ in the denominator of the rhs of Eq.(\ref{P26}).
However, using the Eqs.(\ref{P28})-(\ref{P29}) one can obtain the "pure" solution (but with singular first derivative at $\alpha=\pi/2$) for even values of $k/2$.
Thus, using the Eqs.(\ref{P28})-(\ref{P29}) one obtains the coefficient
\begin{equation}\label{A5}
B_0=\frac{1}{4(k+2)^2},
\end{equation}
which enable one to derive the "pure" physical solution (\ref{P32}) for $k=0,4,8,...$

Application of Eqs.(\ref{A1})-(\ref{A3}) for $l=1$ yields
\begin{equation}\label{A6}
P_\frac{k+1}{2}^\frac{3}{2}(\cos \alpha)=-\sqrt{\frac{2}{\pi \sin \alpha}}
\left\{\cot \alpha\cos\left[\left(\frac{k}{2}+1\right)\alpha\right]+
\left(\frac{k}{2}+1\right)\sin\left[\left(\frac{k}{2}+1\right)\alpha\right]\right\},
\end{equation}
\begin{equation}\label{A7}
Q_\frac{k+1}{2}^\frac{3}{2}(\cos \alpha)=\sqrt{\frac{\pi}{2 \sin \alpha}}
\left\{\cot \alpha\sin\left[\left(\frac{k}{2}+1\right)\alpha\right]-
\left(\frac{k}{2}+1\right)\cos\left[\left(\frac{k}{2}+1\right)\alpha\right]\right\}.
\end{equation}
Inserting representations (\ref{A6}) and (\ref{A7}) into the general formulas (\ref{P16})-(\ref{P18}) with $l=1$, and performing integration, one obtains the particular solution of Eq.(\ref{P10}) in the form
\begin{equation}\label{A8}
f_{k,1}^{(p)}(\alpha)=
\frac{\cos(n\alpha)\left[2n^3+n(n^2-1)\alpha \cot \alpha  \right]+
\sin(n\alpha)\left[n^2(n^2-1)\alpha+(1-3n^2)\cot \alpha \right]}{16n^2(n^2-1)},
\end{equation}
where $n=k/2+1$.
Then, using the result (\ref{A8}) and also Eqs.(\ref{P26}) and (\ref{P27}), one obtains the coefficient
\begin{equation}\label{A9}
A_1=\frac{1}{2\sqrt{2\pi}(k+2)^2}.
\end{equation}
Application of Eq.(\ref{P24}) with using Eqs.(\ref{A8}), (\ref{A9}), (\ref{P13}) and (\ref{A7}) yields the continues (and hence, "pure")  physical solution
\begin{equation}\label{A10}
g_{k,1}(\alpha)=\frac{1}{32 \sin \alpha}\left\{\sin\left[\left(\frac{k+2}{2}\right)\alpha\right]S_{1,\frac{k+2}{2}}(\alpha)+
\cos\left[\left(\frac{k+2}{2}\right)\alpha\right]T_{1,\frac{k+2}{2}}(\alpha)
\right\},
\end{equation}
where the functions $S_{1,n}(\alpha)$ and $T_{1,n}(\alpha)$ being valid only for even values of $k/2\geq2$
are presented in Table \ref{T1}.

For odd values of $k/2\geq1$ we can obtain only the "pure" (but not continues) solution. Using Eqs.(\ref{P28}) and (\ref{P29}), one obtains the coefficient
\begin{equation}\label{A11}
B_1=\frac{3}{4(k+2)^2},
\end{equation}
which enables one to derive the "pure" physical solution of the form (\ref{A10}), where the functions $S_{1,n}(\alpha)$ and $T_{1,n}(\alpha)$ being valid only for odd values of $k/2\geq1$
are presented in Table \ref{T2}.

The methods described in Sec.\ref{S2}, and illustrated in this Appendix on examples of $l=0,1$ can be certainly applied to obtain the physical solution of Eq.(\ref{P8}) with $l>1$.

Thus, Eqs.(\ref{A1})-(\ref{A3}) enable one to obtain the proper representations for the associated Legendre functions (\ref{A2}) with $l>1$. Using these transcendental functions, one can derive the particular solutions of Eq.(\ref{P10}) in the form:
\begin{eqnarray}\label{A12}
f_{k,2}^{(p)}(\alpha)=\frac{1}{128n^2(n^2-4)(n^2-1)\sin^2 \alpha}
~~~~~~~~~~~~~~~~~~~~~~~~~~~~~~~~~~~~~~~~~~~~~~~\nonumber~~~~~\\
\left\{
\sin(n\alpha)\left[(n^2-4)^2(3n^2-1)+3n^2(n^4-5n^2+4)\alpha \sin(2\alpha)-(3n^6+5n^4-34n^2+8)\cos(2\alpha)
\right]
\right.
~\nonumber~~\\
\left.
+\cos(n\alpha)\left[\alpha(n^4-5n^2+4)n\left((n^2+2)\cos(2\alpha)-n^2+4\right)+6(2n^2-5)n^3\sin(2\alpha)
\right]\right\},~~~~~~~~~
\end{eqnarray}

\begin{eqnarray}\label{A13}
f_{k,3}^{(p)}(\alpha)=\frac{1}{768\sin^3 \alpha}
\bigg\{
\sin(n\alpha)\bigg[\alpha \sin \alpha\left[(n^2+11)\cos(2\alpha)-n^2+19\right]
\bigg.\bigg.
~~~~~~~~~~~~~~~~~~~~~\nonumber~\\
\left.
+\frac{3[5( n^2-3)n^2+4](n^2-9)^2\cos \alpha+3(-5n^8+35n^6+21n^4-183n^2+36)\cos(3\alpha)}
{n^2(n^2-9)(n^2-4)(n^2-1)}
\right]
~~~\nonumber~~\\
+\frac{\cos(n\alpha)}{n}\left[
6\alpha\cos \alpha[(n^2+1)\cos(2\alpha)-n^2+4]+\frac{(n^6+43n^4-137n^2-99)\sin(3\alpha)}{2(n^2-1)(n^2-9)}
\right.
~~~~~~\nonumber~\\
\left.\left.
-\sin \alpha\left(\frac{2(3n^2-4)[(n^2+11)\sin^2\alpha-15]}{n^2-4}+ \frac{3(n^2-9)(n^2+1)}{2(n^2-1)}\right)
\right]\right\},~~~~~~~~~~~~~~
\end{eqnarray}
where again $n=k/2+1$.
It should be emphasized that all the particular solutions are correct for any integer $k$ (not necessarily even).

Eqs.(\ref{P26}) and (\ref{P27}) yield the coefficients
\begin{equation}\label{A14}
A_2=-\frac{1}{4\sqrt{2\pi}~k(k+4)},~~~~~~~A_3=\frac{3k^2+12k-4}{48\sqrt{2\pi}~(k-2)(k+2)^2(k+6)},
\end{equation}
which together with the particular solutions (\ref{A12}) and (\ref{A13}) enable one to obtain the continues physical solutions in the form (\ref{P31}) with the function $S_{l,n}(\alpha)$ and $T_{l,n}(\alpha)$ presented in Table \ref{T1} for $l=1,2,3$.

In its turn, using the Eqs.(\ref{P28}) and (\ref{P29}) one obtains the coefficients
\begin{equation}\label{A15}
B_2=\frac{5k(k+4)+16}{4k(k+2)^2(k+4)},~~~~~~~~~~~B_3=\frac{7k(k+4)-20}{4(k-2)(k+2)^2(k+6)},
\end{equation}
which together with the particular solutions (\ref{A12}) and (\ref{A13}) enable one to obtain the "pure" physical solutions in the form (\ref{P31}) with the function $S_{l,n}(\alpha)$ and $T_{l,n}(\alpha)$ presented in Table \ref{T2} for $l=1,2,3$.
Remind that the results presented in Table \ref{T1} are valid for odd values of $(k+2l)/2$, whereas the results presented in Table \ref{T2} are valid for even values of $(k+2l)/2$.

\newpage
\begin{table}
%\begin{center}
\caption{Functions $S_{l,n}(\alpha)$ and $T_{l,n}(\alpha)$ for \textbf{odd} values of $(k+2l)/2=n+l-1$. }
\begin{tabular}{|c|c|}
\hline
 {\normalsize $l$ } & {\normalsize $S_{l,n}{(\alpha)}$}\tabularnewline
\hline
\hline
 $1$ & $2\alpha \sin \alpha-\left(\frac{4 }{n^2-1}\right)\cos \alpha$ \tabularnewline
\hline
 $2$ & $3\alpha\sin(2\alpha)+\frac{n^2-4}{n^2}-\frac{(n^4+10n^2-8)}{n^2(n^2-4)}\cos(2\alpha)$ \tabularnewline
\hline
$3$ & $\frac{\alpha}{2}\left[(n^2+11)\sin(3\alpha)-3(n^2-9)\sin \alpha\right]-
\frac{6(n^4+2n^2-19)}{(n^2-1)(n^2-9)}\cos(3\alpha)+\frac{6(n^2-9)}{n^2-1}\cos \alpha
$ \tabularnewline
\hline\hline
 {\normalsize $l$ } & {\normalsize $T_{l,n}{(\alpha)}$}\tabularnewline
\hline
\hline
 $1$ & $\frac{1}{n}\left[2\alpha \cos \alpha+2\left(\frac{n^2+1}{n^2-1}\right)\sin \alpha\right]$ \tabularnewline
\hline
 $2$ & $\left(\frac{6n}{n^2-4}\right)\sin(2\alpha)-\frac{\alpha}{n}\left[n^2-4-(n^2+2)\cos(2\alpha)\right]$ \tabularnewline
\hline
 $3$ & $\frac{3\alpha}{n}\left[(n^2+1)\cos(3\alpha)-(n^2-9)\cos \alpha\right]+
 \frac{n^6+43n^4-137n^2-99}{2n(n^2-1)(n^2-9)}\sin(3\alpha)-\frac{3(n^2-9)(n^2+1)}{2n(n^2-1)}\sin \alpha$
\tabularnewline
\hline
\end{tabular}
\label{T1}
%\end{center}
\end{table}

\begin{table}
%\begin{center}
\caption{Functions $S_{l,n}(\alpha)$ and $T_{l,n}(\alpha)$ for \textbf{even} values of $(k+2l)/2=n+l-1$. }
\begin{tabular}{|c|c|}
\hline
 {\normalsize $l$ } & {\normalsize $S_{l,n}{(\alpha)}$}\tabularnewline
\hline
\hline
 $1$ & $2\alpha \sin \alpha-\frac{3n^2+1 }{n^2(n^2-1)}\cos \alpha$ \tabularnewline
\hline
 $2$ & $3\alpha\sin(2\alpha)+\frac{n^2-4}{2n^2}-\frac{(n^4+22n^2-8)}{2n^2(n^2-4)}\cos(2\alpha)$ \tabularnewline
\hline
$3$ & $\frac{\alpha}{2}\left[(n^2+11)\sin(3\alpha)-3(n^2-9)\sin \alpha\right]+
\frac{3}{2n^2(n^2-1)}
\left[(n^2-9)(3n^2+1)\cos \alpha-\frac{(n^2-3)(3n^4+26n^2+3)}{n^2-9}\cos(3\alpha)\right]
$ \tabularnewline
\hline\hline
 {\normalsize $l$ } & {\normalsize $T_{l,n}{(\alpha)}$}\tabularnewline
\hline
\hline
 $1$ & $\frac{1}{n}\left[2\alpha\cos \alpha+\left(\frac{n^2+3}{n^2-1}\right)\sin \alpha\right]$ \tabularnewline
\hline
 $2$ & $\frac{3(3n^2+4)}{2n(n^2-4)}\sin(2\alpha)-\frac{\alpha}{n}\left[n^2-4-(n^2+2)\cos(2\alpha)\right]$ \tabularnewline
\hline
 $3$ & $\frac{3\alpha}{n}\left[(n^2+1)\cos(3\alpha)-(n^2-9)\cos \alpha\right]+
 \frac{(n^6+85n^4-173n^2-297)}{4n(n^2-1)(n^2-9)}\sin(3\alpha)-
 \frac{3(n^2-9)(n^2+3)}{4n(n^2-1)}\sin \alpha
  $
\tabularnewline
\hline
\end{tabular}
\label{T2}
%\end{center}
\end{table}

\end{document}